\documentstyle[12pt,moriond,psfig]{article}

\def\spose#1{\hbox to 0pt{#1\hss}}

\def\kms{\ifmmode {\rm\,km\,s^{-1}}\else ${\rm\,km\,s^{-1}}$\fi}

\def\kmsmpc{\ifmmode {\rm\,km\,s^{-1}\,Mpc^{-1}}\else ${\rm\,km\,s^{-1}\,Mpc^{-1}}$\fi}

\def\ergps{\ifmmode {\rm\,erg\,s^{-1}}\else ${\rm\,erg\,s^{-1}}$\fi}
\def\ergpscm2{\ifmmode {\rm\,erg\,s^{-1}\,cm^{-2}}\else
    ${\rm\,erg\,s^{-1}\,cm^{-2}}$\fi}

\def\deg{\ifmmode {^{\circ}}\else {$^\circ$}\fi}
\def\degr{\ifmmode {^{\circ}}\else {$^\circ$}\fi}
\def\degs{\ifmmode {^{\circ}}\else {$^\circ$}\fi}

\def\h3Mpc{h^{-3}{\rm Mpc}^3}

\def\arcsec{\ifmmode {^{\prime\prime}}\else $^{\prime\prime}$\fi}
\def\asec{\ifmmode {^{\prime\prime}}\else $^{\prime\prime}$\fi}
\def\arcmin{\ifmmode {^{\prime}}\else $^{\prime}$\fi}
\def\amin{\ifmmode {^{\prime}}\else $^{\prime}$\fi}

\def\secper{\ifmmode \rlap.{^{s}}\else $\rlap{.}{^{s}} $\fi}
\def\minper{\ifmmode \rlap.{^{m}}\else $\rlap{.}{^m} $\fi}
\def\secspt{\ifmmode \rlap.{^{\prime\prime}}\else
    $\rlap.{^{\prime\prime}}$\fi}
\def\arcsper{\ifmmode \rlap.{^{\prime\prime}}\else
    $\rlap.{^{\prime\prime}}$\fi}
\def\minspt{\ifmmode \rlap.{^{\prime}}\else
    $\rlap.{^{\prime}}$\fi}
\def\arcmper{\ifmmode \rlap.{^{\prime}}\else
    $\rlap.{^{\prime}}$\fi}
\def\spose#1{\hbox to 0pt{#1\hss}}
\def\simlt{\mathrel{\spose{\lower 3pt\hbox{$\mathchar"218$}}
     \raise 2.0pt\hbox{$\mathchar"13C$}}}
\def\simgt{\mathrel{\spose{\lower 3pt\hbox{$\mathchar"218$}}
     \raise 2.0pt\hbox{$\mathchar"13E$}}}

\def\U300{\ifmmode{U_{300}}\else{$U_{300}$}\fi}
\def\B450{\ifmmode{B_{450}}\else{$B_{450}$}\fi}
\def\V606{\ifmmode{V_{606}}\else{$V_{606}$}\fi}
\def\I814{\ifmmode{I_{814}}\else{$I_{814}$}\fi}
\def\J110{\ifmmode{J_{110}}\else{$J_{110}$}\fi}
\def\H160{\ifmmode{H_{160}}\else{$H_{160}$}\fi}

\begin{document}

\heading{GALAXY EVOLUTION AT \mbox{\boldmath $0 < z < 2$} FROM THE NICMOS HDF--NORTH}
 
\author{Mark Dickinson} {STScI}{Baltimore, USA}

\begin{moriondabstract}

We have carried out a deep infrared imaging survey (1.1$\mu$m and 
1.6$\mu$m) of the Hubble Deep Field North (HDF--N) using NICMOS 
on board the {\it Hubble Space Telescope.}   The combined 
WFPC2+NICMOS data set lets us study galaxy morphologies, colors and 
luminosities at common rest frame wavelengths over a broad range 
of redshifts, e.g., in the $V$--band out to $z = 2$.  Here, 
I illustrate some applications of this data set for studying 
the evolution of giant galaxies, on and off the Hubble Sequence.
Large, relatively ordinary spiral galaxies are found out to at 
least $z \approx 1.25$.  Morphological irregularities seen in many 
distant HDF galaxies tend to persist from ultraviolet through optical 
rest frame wavelengths, suggesting that these are genuinely peculiar,
structurally disturbed systems.  Red giant ellipticals are found 
out to (photometric) redshifts $z \approx 1.8$, implying that some 
such galaxies probably formed the bulk of their stars at $z_f \simgt 4$.
However, there are also bluer early type galaxies at $z > 0.5$, 
which may have experienced extended star formation histories.  
Finally, there appears to be a substantial deficit of high luminosity 
galaxies of all types at $1.4 \simlt z < 2$ compared to lower 
redshifts.  However, this result must be considered with caution 
given the small volume of the HDF, its susceptibility to 
line--of--sight clustering variations, and the heavy reliance 
on photometric redshifts at $z \simgt 1.4$.

\end{moriondabstract}
 
\section{Introduction}

The Hubble Deep Fields (North and South, or HDF--N and HDF--S)
currently offer the deepest optical images of the distant universe, 
and their exquisite angular resolution provides the opportunity to 
study the morphologies of galaxies in detail and explore how the 
galaxy population has transformed with cosmic time.  When considering 
the properties of HDF galaxies, it is important to keep some 
facts/limitations of the data set in mind.  The co--moving volume 
probed by the central, deepest WFPC2 fields (i.e., neglecting the 
shallower but wider flanking fields) is quite small.  
For the currently popular ``supernovae + Cepheids'' 
cosmology, i.e., $\Omega_M = 0.3$, $\Omega_\Lambda = 0.7$, 
$H_0 = 70$~km/s/Mpc, which I will adopt here,
the comoving volumes out to $z = 1$ and 2 are approximately 5000 
and 20000 Mpc$^3$, respectively.  Multiplying these by the 
normalization of the local galaxy luminosity function (using 
$\phi^\ast = 0.0055h_{70}^3$~Mpc$^{-3}$, from \cite{Gardner 1997})
gives a rough estimate of the number of $L^\ast$ galaxies 
(or their progenitors) expected within the HDF volume.  This is
only $\sim 30$ at $z < 1$.  Thus, small number statistics alone 
limit the utility of the HDF for a reliable census of bright galaxy 
properties at $z < 1$, and given real galaxy clustering the possible 
uncertainties are greater still.  

At $1 < z < 2$ there is room enough for $\sim 80$ $L^\ast$ galaxies, 
still small but much better for statistical purposes.  However, at 
$z > 1$, the optical rest frame, where we are most familiar with local 
galaxy properties, redshifts into the near--infrared.  WFPC2 imaging 
therefore measures the rest frame ultraviolet properties of many 
(probably most) faint HDF galaxies, and thus primarily traces the 
light of hot, short--lived stars, modulated by the possible effects 
of extinction.  Systematically comparison of $z > 1$ HDF galaxies to those 
at $z < 1$ therefore requires deep, near--infrared data.  The HDF--N was 
observed in the near--infrared from the ground in several different programs 
(\cite{Hogg 1997}, \cite{Barger 1998}, \cite{Dickinson 1998}).
The depth and angular resolution (typically $\sim 1\arcsec$) of these 
data are a poor match to that of the optical WFPC2 HDF images.  Two 
programs therefore targeted the HDF--N with NICMOS on board HST, providing 
much deeper images with high angular resolution.   The NICMOS GTOs 
\cite{Thompson 1999} imaged one NIC3 field ($\sim 51\arcsec\times 51\arcsec$) 
for 49 orbits each at F110W (1.1$\mu$m) and F160W (1.6$\mu$m).  We, on
the other hand, took the ``wide field'' approach, mosaicing the complete 
HDF--N with a mean exposure time of 12600s per filter in F110W and F160W.
Sensitivity varies over the field of view, but the mean depth is 
$AB \approx 26.1$ at $S/N=10$ in an $0\secspt7$ diameter aperture.  
The drizzled PSF has FWHM~=~$0\secspt22$, primarily limited by the 
NIC3 pixel scale.   

The NICMOS data, then, offer the opportunity to study the photometric 
and morphological properties of HDF galaxies at rest frame optical 
wavelengths out to $z \approx 3$.  The $H_{160}$ bandpass\footnote{I will 
use AB magnitudes here throughout, and notate the six WFPC2+NICMOS 
bandpasses by \U300, \B450, \V606, \I814, \J110 and \H160.} samples 
the rest frame $I$, $V$, and $B$ bands at $z \approx 1$, 2, and 2.8, 
respectively.  The combined WFPC2+NICMOS data set offers several 
attractive new options for studying distant galaxies.  Given the 
redshift (or a photometric estimate thereof) for an object, we can 
measure magnitudes and colors at fixed {\it rest} frame wavelengths, 
and even generate fixed rest frame {\it images} to study galaxy 
morphologies in a common manner over a very broad range of redshifts.
I will try to demonstrate here how this can be quite an educational 
way to study distant galaxies.

For this review, I will discuss the morphologies, colors, and space 
densities of giant galaxies in the HDF, on and off the Hubble sequence, 
at $0 < z \simlt 2$, where the NICMOS images probe rest frame optical 
wavelengths at the rest frame $V$ band or redder.  The spectroscopic 
redshift sample (primarily from \cite{Cohen 1996} and \cite{Phillips 1997}) 
becomes thin (nil, in fact) at $1.4 < z < 2$, and I will rely on 
photometric redshifts fit to our 7--band WFPC2+NICMOS+$K_s$ data by 
Tamas Budav\'ari and collaborators (\cite{Budavari 1999}, \cite{Csabai 1999}).
The near--infrared photometry should substantially improve the reliability 
of photometric redshifts (\cite{Connolly 1997}, \cite{Fernandez-Soto 1999}), 
but it is nevertheless important to remember that there are, as yet, 
no spectroscopic calibrators for the $z_{phot}$ estimates in this 
redshift range.   I will restrict the $z_{phot}$ sample to $\H160 < 26$, 
where we believe the photometry, completeness and reliability to be good.  

\section{Spiral and irregular galaxies}

WFPC2 imaging established the apparent overabundance of distant, 
morphologically irregular galaxies falling outside the standard Hubble 
sequence ``tuning fork.''    Galaxy number counts divided by morphological 
type from the Medium Deep Survey (\cite{Casertano 1995}, \cite{Driver 1995}, 
\cite{Glazebrook 1995}) found spiral and elliptical galaxies in numbers 
roughly comparable to predictions from no--evolution or pure luminosity 
evolution (PLE) models.  However, irregular galaxies were far more common 
than expected, suggesting that they are primarily responsible for 
the faint blue galaxy excess.  In the HDF, $\sim$40\% of galaxies at 
$\I814 = 25$ fall into the irregular/peculiar/merging category 
\cite{Abraham 1996}.  WFPC2 images, however, did not resolve the 
question of how much the so--called ``morphological $k$--correction'' 
might influence these assessments.  At $R = 24$, the median redshift 
in the HDF~\cite{Cohen 2000} is $\langle z \rangle = 1$, and at 
$I = 25$ a substantial majority of objects should be at $z \simgt 1$,
where WFPC2 samples rest frame ultraviolet light.  Thus the observed 
morphologies could primarily map the distribution of young, UV--bright 
star forming regions and the obscuring effects of dust rather than
the overall structure of the stellar mass.  Early efforts to simulate 
the appearance of high redshift galaxies by artificially redshifting 
vacuum UV images of nearby galaxies demonstrated both that irregular 
morphologies might be expected, and that the lower surface brightness 
features of ``normal'' galaxies today might be hard to see at high 
redshift due to $(1+z)^4$ dimming (\cite{Bohlin 1991},
\cite{Giavalisco 1996}, \cite{Hibbard 1997}).

\begin{figure}
\centerline{\psfig{figure=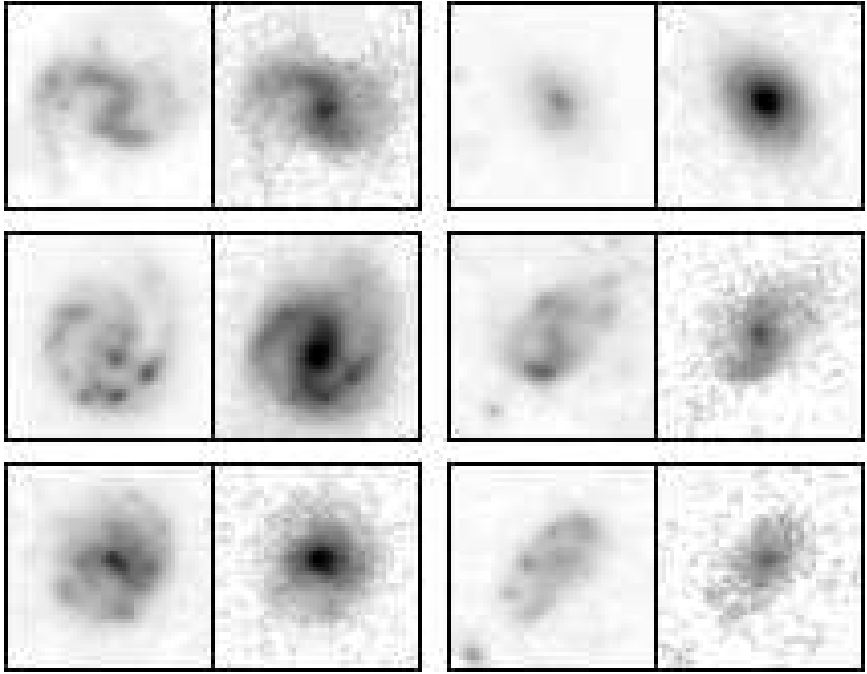,width=4.5in}}
\caption{Examples of large disk galaxies at $0.96 < z < 1.23$
from the HDF--N.  For each pair of images, the left panel
shows the galaxy in the UV at rest frame 3000\AA, while the
right shows the rest frame $R$--band at 6500\AA.  Boxes
are $4\arcsec \times 4\arcsec$, corresponding to 
32$h_{70}^{-1}$~kpc on a side at $z=1$;  neighboring objects
have been masked out.   Bulges, bars, and interarm disk light 
are more prominent at the longer wavelengths, as expected, and are 
sometimes wholly invisible in the WFPC2 data.}
\end{figure}

Such questions can largely be resolved by directly examining how 
the morphology of distant galaxies changes with wavelength.  The 
WFPC2+NICMOS HDF data set allows us to form images of galaxies at 
fixed rest frame wavelengths over a wide range of redshifts.  
Figure~1 illustrates several $z \sim 1$ disk galaxies, interpolating 
between bandpasses to rest frame wavelengths 3000\AA\ 
and 6500\AA.  Here, the morphological differences are much as one would 
expect:  the spiral arms and HII regions are prominent in the UV rest 
frame images.  In the rest frame $R$--band, the inter--arm disk light 
is stronger, the spiral arms tend to regularize, and prominent bulges 
and bars often appear that are all but invisible in the WFPC2 images.

\begin{figure}
\centerline{\psfig{figure=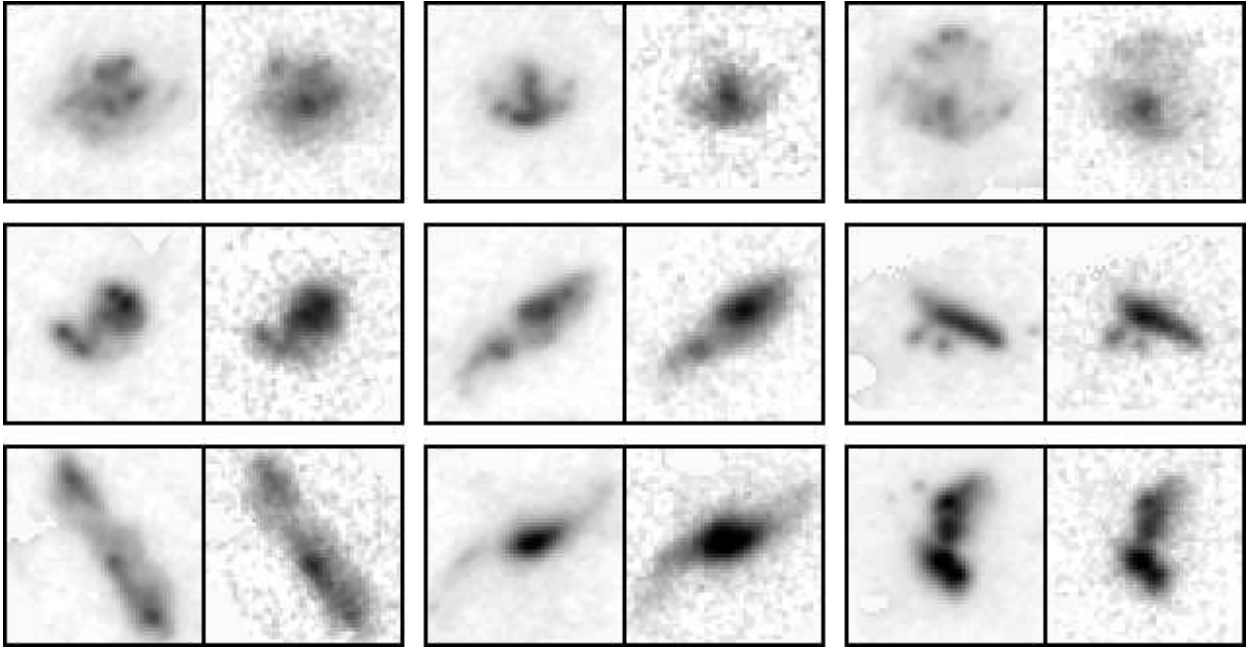,width=6.5in}}
\caption{HDF--N objects with irregular morphologies at
$0.75 < z < 1.36$, at rest frame 3000\AA\ (left) and 
6500\AA\ (right).  Again, box sizes are $\sim 32h_{70}^{-1}$~kpc.
These galaxies are large and quite luminous by comparison 
to typical irregular galaxies in the local universe.  
Although some of the galaxies are more centrally concentrated 
at longer wavelengths, in general the peculiar morphologies 
are preserved from over long wavelength baselines.}
\end{figure}

Some examples of galaxies with peculiar WFPC2 morphologies are
shown in Figure~2.  The irregularities in these objects tend to be 
preserved across the UV--to--optical wavelength baseline:  dramatic 
transformations, where peculiar objects are revealed to be comparatively 
ordinary galaxies at longer wavelengths, are comparatively rare.  
The structure of ``chain galaxies'' like 2-736.1 ($z = 1.355$,
lower right in Figure~2) is almost entirely unchanged from 1300\AA\
to 6800\AA\ in the rest frame.  Some authors (e.g., \cite{Cowie 1995})
have noted that these structures were unlikely to be stable and 
persist for long given the nominal dynamical time scales for these 
galaxies, and suggested that therefore they must be inherently 
young objects.  For very blue objects like 2-736.1 this is quite 
likely true:  the observed light from the UV through the IR is 
apparently dominated by the same, relatively young generation 
of stars, and if there is an older stellar component it is either 
well mixed with the younger stars, or its light is entirely 
swamped by the dominant, younger population out to long 
wavelengths.  Some of the irregular galaxies appear to be
simply very late type disks, with weak or absent bulges
and without well--ordered spiral structure.  There are also 
{\it red,} asymmetric galaxies \cite{Conselice 2000}, a comparative 
rarity in the local universe:  these may be objects where recent 
encounters or mergers have disturbed the morphologies without inducing
much star formation, or in some cases dust--reddened systems.

From a preliminary analysis of structural parameters for the overall 
HDF population at $H_{160} \simlt 24$, we find \cite{Conselice 2000} 
that galaxies in the NICMOS images tend to have smaller half--light 
radii, to be more centrally concentrated, and to exhibit greater symmetry 
in the the WFPC2 data.  At fainter magnitudes, any trends are more difficult 
to discern because the mean galaxy size in the NICMOS data becomes small 
enough that the PSF dominates structural measurements.  Other
studies \cite{Teplitz 1998} have noted similar trends in 
among galaxies in NICMOS parallel images, although few fields had 
both optical and infrared HST imaging to permit direct 
comparisons on a galaxy--by--galaxy basis.  In general, it appears 
that the morphological peculiarities seen in deep WFPC2 images arise 
from a variety of effects.  Giant spiral and elliptical galaxies 
are present out to at least $z=1.3$ and perhaps beyond (see below), 
and look comparatively normal in the NICMOS data.   But in general, 
the impression that the distant universe is rich in irregular and 
disturbed objects is preserved when considering the NICMOS images.

\section{Elliptical galaxies}

The myth of the passively evolving elliptical galaxy, formed 
{\it in situ} at high redshift in a dissipationless collapse 
and starburst, then burning its main sequence away for billions 
of years thereafter, has held sway since the scenario was 
postulated \cite{ELS} and its photometric consequences were first 
modeled (\cite{Larson 1974}, \cite{Tinsley 1976}).  The broad homogeneity 
of giant elliptical galaxy photometric and structural properties 
in the nearby universe has compelled many investigators to hold this 
``monolithic'' formation scenario as a rare example of a clearly stated 
null hypothesis for galaxy evolution against which to compare detailed 
measurements and computations.  With HST imaging, we can directly 
observe the evolutionary history of the elliptical galaxy population.  
Until recently, much of this work has been done 
using rich clusters, where early type galaxies dominate the population.
I will not review the cluster work here, except to note that most 
observers have favored the broad interpretation of quiescent, 
nearly passive evolution among cluster ellipticals out to 
$z \approx 1$.\footnote{The introduction to \cite{Schade 1999} 
provides a recent, succinct and comprehensive review of the literature 
concerning elliptical galaxy evolution at $0 < z < 1$, both in clusters 
and in the field.}

It is more challenging to uniformly select and study samples of high 
redshift {\it field} ellipticals.  Attempts to do so have variously 
used selection criteria based on morphology, color, or both, and only 
a few have incorporated redshift information.  There is little evolution 
in the luminosity function of intrinsically red galaxies from the 
CFRS \cite{Lilly 1995}, but this appears to contradict basic expectations 
from PLE models, where galaxies should be brighter at higher redshift 
\cite{Kauffmann 1996}.  This might imply that elliptical galaxies 
assemble late by merging processes, or that some fraction of distant 
ellipticals are blue enough to drop out of color--selected samples.  
Indeed, morphologically defined samples from HST imaging (e.g., 
\cite{Schade 1999}, \cite{Kodama 1999}) have identified bluer field 
ellipticals which might account for the decline in number density
at higher redshifts seen in the color--selected samples.

At $z > 1$, the strong $k$--correction for early--type galaxies 
means that infrared data are required to take an unbiased census.  
In the HDF, studies using ground--based infrared data have found
a deficit of red ellipticals at $z > 1$ (\cite{Zepf 1997}, 
\cite{Franceschini 1998}, \cite{Barger 1998}), although other
infrared surveys have found higher surface densities of red galaxies 
(\cite{McCracken 1999}, \cite{Eisenhardt 1998}), raising concerns 
about field--to--field variations.   The most extensive optical--infrared 
color surveys incorporating WFPC2 \cite{Menanteau 1999} 
or NICMOS \cite{Treu 1999} morphologies have also concluded that there 
are fewer bright, red ellipticals at $z > 1$ than would be expected 
from PLE models.  Individual examples 
of red, high redshift ellipticals have been found in deep HST images 
(e.g., the HDF--S NICMOS field \cite{Stiavelli 1999}, \cite{Benitez 1999}),
although very few such galaxies have spectroscopic redshifts
(see \cite{Dunlop 1996}, \cite{Spinrad 1997}, \cite{Soifer 1999} 
for rare examples).

\begin{figure}
\centerline{\psfig{figure=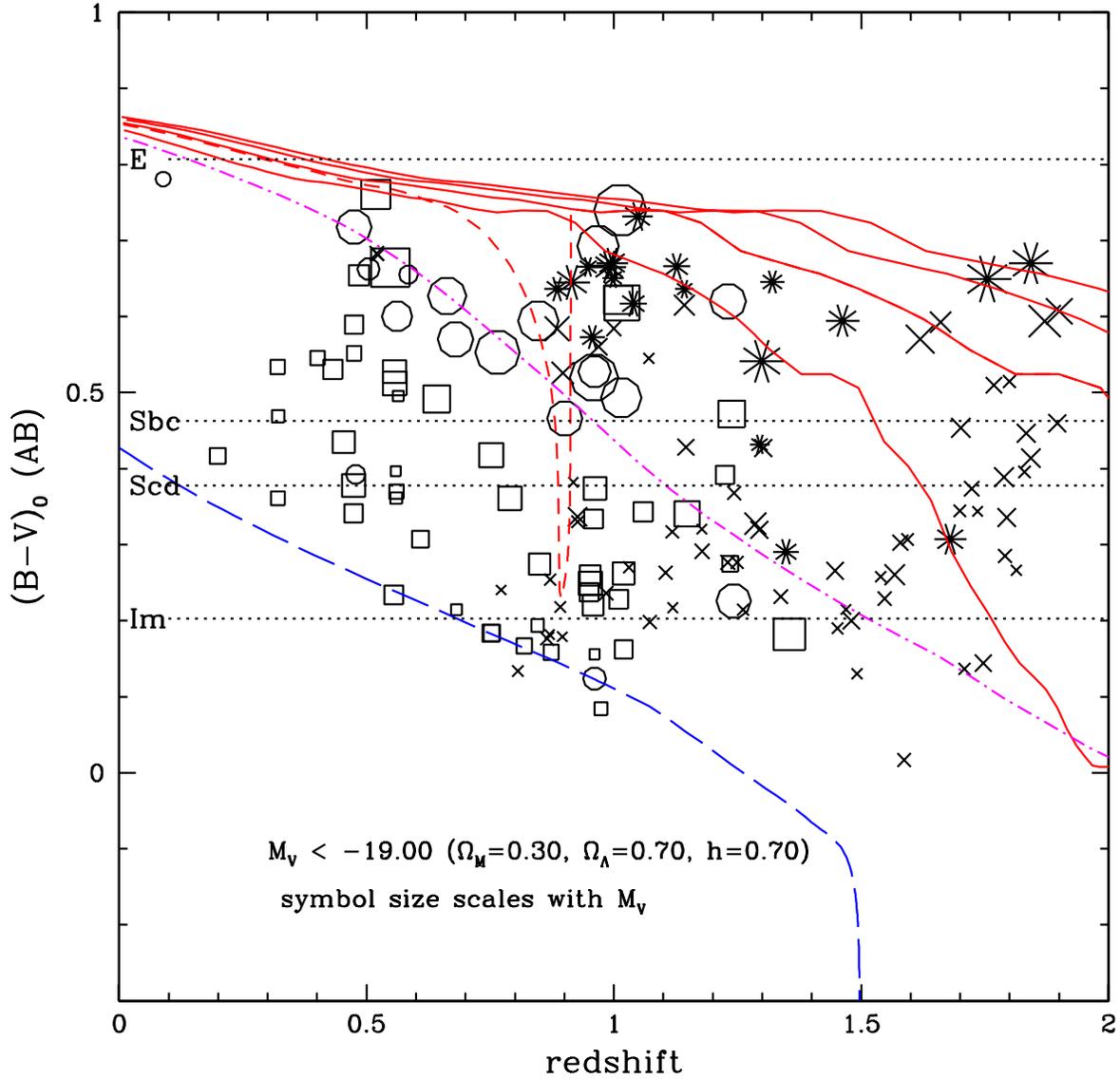,width=6.5in}}
\caption{Rest frame $B-V$ colors for a sample of HDF galaxies with
$M_V < -19$.  Open and skeletal symbols indicate galaxies with 
spectroscopic and photometric redshifts, respectively.  Circles
and ``asterisks'' indicate galaxies which we have classified
morphologically as ``early type'' (E and S0).  Typical $z=0$
colors of galaxies derived from empirical spectral templates
\cite{CWW} are indicated at left.  Lines indicate solar 
metallicity population synthesis \cite{BC96} color evolution 
models.  The solid lines show 0.1~Gyr burst models with formation 
redshifts $z_f = 5$, 4, 3 and 2.1 (from red to blue).  Ellipticals 
with red colors consistent with high formation redshifts are found out 
to $z \approx 1.8$.  However, bluer ellipticals are also found 
at $0.5 \simlt z \simlt 2$, with colors plausibly matched by models 
with extended star formation rate (SFR) histories (dot--dashed line:  
$z_f = 2.5$, $\tau = 1$~Gyr exponential SFR), or by late bursts on 
an otherwise old galaxy (short--dashed line:  2\% burst by mass 
superimposed on a $z_f = 4$ passive model).  The long--dashed line 
is a $\tau = 5$~Gyr exponential SF model with $z_f = 1.5$ that 
would match the colors of present--day mid--type spirals.}
\end{figure}

Figure~3 shows the rest frame $(B-V)_0$ colors of HDF galaxies out to
$z = 2$, derived by interpolation between observed bandpasses to the fixed 
rest frame wavelengths.  Spectroscopic redshifts are used wherever possible,
and photometric redshifts otherwise.  A ``volume limited'' sample of 
galaxies is plotted, selected to have $M_V < -19$ at all redshifts.
The galaxies mostly span a range of $0.1 < (B-V)_0 < 0.8$ at all redshifts,
as do present--day galaxies.
At $z \simgt 0.5$, an increasing fraction of galaxies bluer than 
present--day Scd spirals are found.

Using a combination of visual classifications, surface brightness 
profile fitting, and concentration/asymmetry measurements, we have 
defined a subsample of morphologically selected ``early type'' galaxies 
(roughly T--types -7 to -2), which are indicated by circle and asterisk 
symbols in the plots.  The ``red envelope'' of the color distribution 
is largely defined by early type galaxies, and becomes gradually bluer 
at higher redshifts, as would be expected from passively evolving 
models.  Galaxies with colors consistent with purely passive 
evolution and a high formation redshift are found out to (photometric) 
redshifts $z \approx 1.8$.  Taking the spectral synthesis models at 
face value, these most distant ellipticals must have formed the bulk 
of their stars at $z \simgt 4$.  They are quite luminous:  with passive 
fading and no further merging or star formation, they would become 
$\sim L^\ast$ galaxies by $z=0$ (see Figure~4).  The $z=1.01$ giant 
elliptical HDF 4--752, perhaps the most intrinsically luminous/massive 
galaxy in the HDF, would fade to $M_V \approx -22 + 5 \log h_{70}$ today.
Curiously, there are few good NICMOS--selected candidates for red 
ellipticals in the HDF at $z > 2$;  perhaps the only one is the 
so--called ``J dropout'' object \cite{Dickinson 2000}, 
which might conceivably be a maximally old elliptical at 
$z \approx 3$ to 4.

At the same time, there are many galaxies which we have classified
as ``ellipticals'' which are substantially bluer than the PLE
predictions,  particularly at $z > 0.5$.  This has also been 
noted previously from investigations of the CFRS+LDSS sample and 
the HDF itself (\cite{Schade 1999}, \cite{Kodama 1999}).  Metallicity 
variations may account for part of the range of colors, but many
of these ``blue ellipticals'' appear to be genuinely outliers 
from the color--magnitude relation.  The rms scatter in $(B-V)_0$
colors at $0.8 < z < 1.1$ is approximately twice that seen among
present--day ellipticals, even when the most extreme outliers
are excluded.   This is quite different than the situation found
in rich cluster environments at similar redshifts \cite{SED98}.
The bluer colors can easily be accommodated with trace amounts 
of later star formation in otherwise old galaxies (see Figure~3).

\section{Galaxy space densities}

Some authors, using infrared--selected galaxy redshift surveys,
have suggested that there are not enough bright galaxies at $z \simgt 1$ 
to account for the present--day population if galaxy number density 
is conserved with redshift (e.g., \cite{Kauffmann 1998}).  The small 
volume of the HDF makes it less than ideal for studying this question, 
but with a deep NICMOS--selected sample we may at least address 
it in the most broad--brush manner.  Figure~4 shows rest frame 
$V$--band luminosities for HDF galaxies at $0 < z < 2$, plotted 
against co--moving volume $V(<z)$.  In such a plot, a constant 
density of points represents a constant co--moving space 
density (e.g., \cite{Cohen 2000}), making it easy to 
``see'' trends in the evolution of the luminosity 
function.    Models for luminosity evolution given various
possible star formation histories are indicated.  If galaxies 
evolved according to such models, then counting objects between 
the parallel tracks would measure their space density with 
redshift.    Let us take the simple exercise of dividing the
volume out to $z=2$ in half:  the midpoint is at $z=1.37$
for the adopted cosmology.  Consider the $z_f = 5$ single 
burst models (solid lines in Figure~4):  the bottom--most
of these lines corresponds to an object with luminosity 
$\approx 0.2 L_V^\ast$ today, and apparent magnitude 
$\H160 = 24$ at $z = 2$.  Our morphological classifications
should be complete at all redshifts brighter than this line.
There are 84 HDF galaxies at $0 < z < 2$ more luminous than 
this model:  the number ratio between the ``low--$z$'' and 
``high--$z$'' volumes is 76:9.  Considering only the morphological 
early type galaxies, the ratio is 27:4, similar to the population 
as a whole.  Of course ``single burst'' models cannot describe
the real history of most galaxies, but the same situation
holds for virtually any scenario in which galaxy number is
conserved.  Even if we were to assume a {\it negative}
luminosity evolution model where galaxies are less
luminous at high redshift (e.g., constant SFR histories,
the dotted lines in Figure~4), the same result holds.
Considering the bottom--most of the constant SFR models,
the low-- to high--$z$ number ratio is 66:21.  

\begin{figure}
\centerline{\psfig{figure=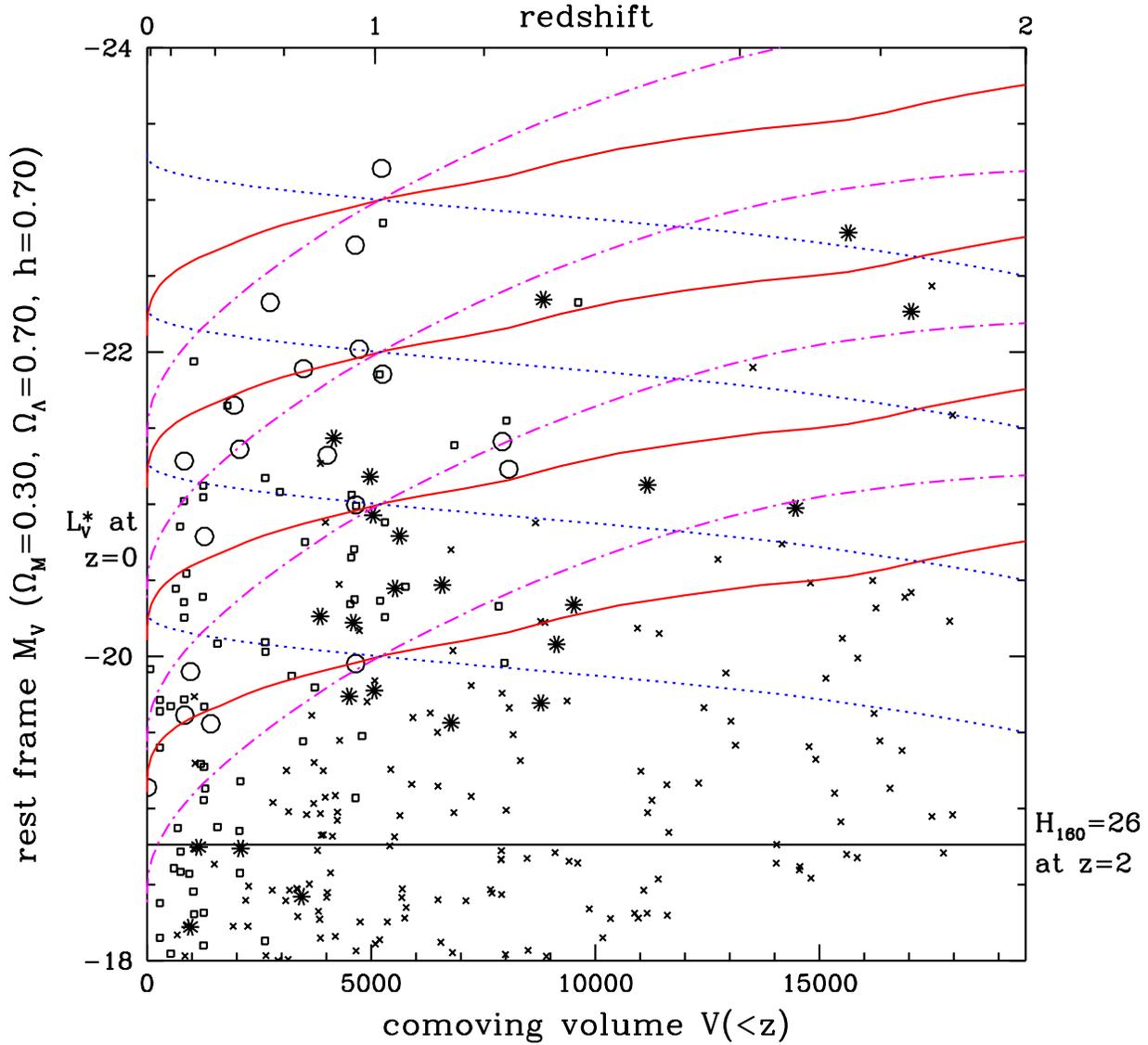,width=6.5in}}
\caption{Rest frame $V$--band absolute magnitude of HDF galaxies
plotted versus co--moving volume out to redshift $z$.  
Symbols are coded as in Figure~3.  The horizontal line near
the bottom marks the $M_V$ corresponding to $H_{160} = 26$
at $z=2$:  the sample should be complete at all redshifts
$0 < z < 2$ brighter than this line (neglecting surface
brightness biases), although morphological classifications
have only been made to $H_{160} = 24$.  Present--day 
$\sim L_V^\ast$ is marked at left.  The curved lines indicate 
luminosity evolution models for various SFR histories, spaced 
by 1~magnitude intervals.  Solid lines:  $z_f = 5$ single burst
models;  Dotted lines:  $z_f = 2.5$ constant SFR models;
Dot--dashed lines:  $z_f = 2.5$, $\tau = 1$~Gyr exponential
SFR models.}
\end{figure}

Detection and photometry biases due to cosmological surface
brightness dimming may contribute to this apparent high 
redshift deficit, but are unlikely to be the dominant effect.
We have carried out simple simulations, taking bright 
($L_V \simgt L^\ast$) HDF galaxies at $z \simlt 1$,
artificially redshifting them to $1.4 < z < 2$ {\it without}
any luminosity evolution, and re--inserting them into the 
NICMOS images at common rest frame wavelengths.  Most
would still be easily detectable;  their recovered magnitudes
can be somewhat biased, although careful choice
of photometric procedures (e.g., using constant metric 
apertures or ``Kron''--style photometry based on moments
of the light profile) should minimize the impact of this
effect.

Several important caveats must be kept in mind, however.
First, there are {\it no} HDF galaxies with spectroscopic
redshifts $1.37 < z < 2$:  this is exactly the ``redshift
desert'' where spectroscopy is most difficult, and thus
our comparison depends entirely on the reliability of our 
photometric redshifts.  Although we believe they are 
good, there could conceivably be some sort of systematic 
``depopulation'' of this uncalibrated region.  This, however,
would have to be a dramatic effect to account for the 
difference:  only if all galaxies without spectroscopic redshifts
and with $0.5 < z_{phot} < 1.37$ were instead assigned 
$z \approx 1.7$ would the number densities balance out,
and photometric redshifts at $z \simlt 1.2$ have 
been shown to be generally quite reliable \cite{Hogg 1998}.
Objects assigned $z_{phot} > 2$ might also be at lower
redshifts, although in general the Lyman break signature makes
such redshift estimates fairly robust.  Second, the HDF--N is 
only a single sight line, and large--scale structure may 
affect results to a much greater degree than Poisson 
statistics.  E.g., there are substantial overdensities in 
the HDF redshift distribution at $z \approx 0.56$, 
0.96, and 1.02 \cite{Cohen 2000}.  The latter two are 
rich in early type galaxies, and indeed there are {\it too many}
bright ellipticals at $z \sim 1$ in the HDF compared to
extrapolations from the present--day luminosity function
or by comparison with other faint field surveys 
(\cite{Schade 1999}, \cite{Stanford 2000}).   
In fact, $\sim$23\% of the rest frame
5400\AA\ luminosity density in the HDF at $z < 1.1$ comes from
just four galaxies (three of which are ellipticals) in the redshift 
spikes at $z=0.96$ and 1.02.   These overdensities appear to be
``walls'' or ``sheets'' whose transverse sizes substantially 
exceed the WFPC2 field of view.  Similar structures are 
now known to be ubiquitous at $z \sim 3$ (\cite{Steidel 1998}, 
\cite{Adelberger 1998}).   Although I have split the HDF volume 
out to $z = 2$ evenly at $z = 1.37$, the line--of--sight 
co--moving path length intervals are very different, 
roughly 4:1.   The number of bright galaxies, even in 
these seemingly broad redshift intervals, might 
just indicate the luck of the draw with encountering 
the most overdense redshift ``spikes.''

The deficit of bright, high redshift galaxies in the HDF--N seems 
to apply not only to ellipticals (\cite{Zepf 1997}, \cite{Franceschini 1998}) 
but to all galaxy types.   Spectroscopic verification of the photometric 
redshifts in this redshift range is clearly critical.  Only similar analyses 
of other sight lines will tell whether this is a universal situation or 
whether it is a feature of this particular sight line.  Although the quality 
and depth of the combined optical+infrared data for the HDF--N is 
unmatched elsewhere, photometric redshift analyses of other ground--based 
and HST deep survey fields have also suggested bright galaxy deficits 
at $z > 1$ \cite{Fontana 1999}.   At the same time, it is notable 
that even at $z \approx 1.8$, the most luminous HDF
galaxies are evidently mature giant elliptical galaxies, with
SEDs that suggest large formation redshifts ($z_f \simgt 4$),
and which even with passive luminosity evolution would fade 
to $L \simgt L^\ast$ objects today.    The 20 brightest galaxies
with $1.37 \simlt z_{phot} \simlt 2$ are shown in Figure 6.
The red ellipticals are among the brightest objects.  Several
appear to be disk galaxies, although few seem to be as large
as the comparably bright $z \sim 1$ HDF spirals shown in Figure~1,
or to have prominent, high surface brightness spiral 
arms.\footnote{In a few cases from Figure~5, spiral structure 
is seen more clearly in the rest--frame UV WFPC2 data, which has 
better angular resolution.}  However, a more careful analysis 
accounting fully for surface brightness dimming is needed before 
reaching firm conclusions.

\begin{figure}
\centerline{\psfig{figure=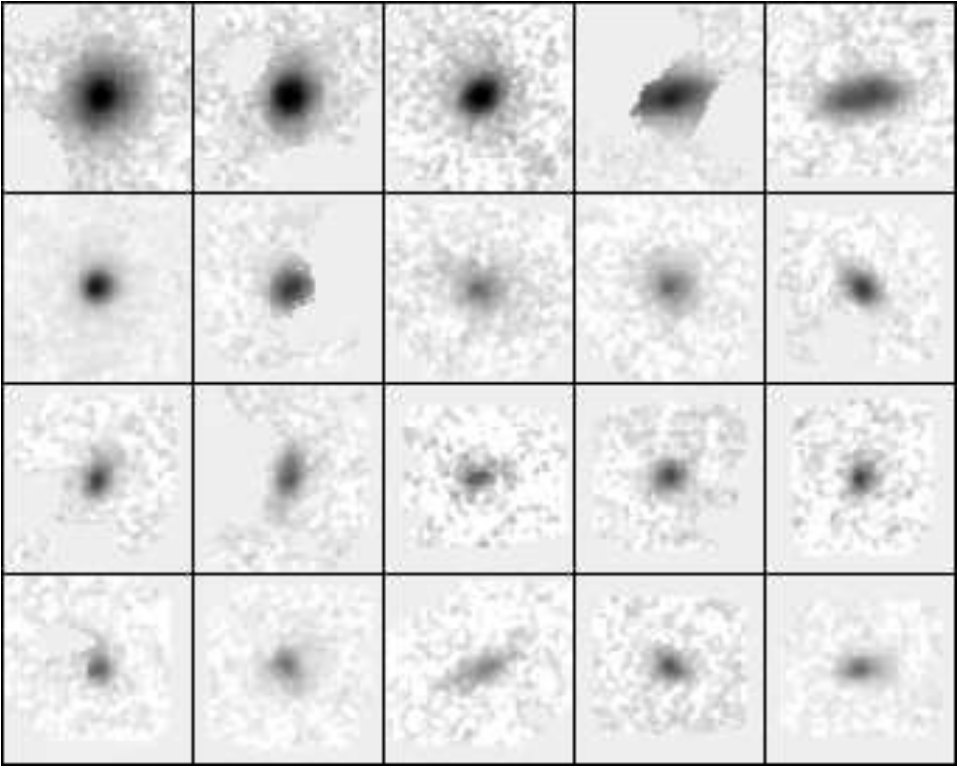,width=5.0in}}
\caption{Rest frame $V$--band images of the 20 most luminous 
HDF galaxies ($M_V < -20$) with $1.37 < z_{phot} < 2$, scaled to 
common physical size and surface brightness and ordered by
luminosity.  The box size is 32$h_{70}^{-1}$~kpc, as 
for Figures 1 and 2.}
\end{figure}

\section{Conclusions}

Deep, high resolution optical--infrared imaging, together with 
spectroscopy and (where needed) photometric redshifts, offers the 
means for studying galaxy properties at common rest frame wavelengths 
over a broad redshift baseline.  At present, the HDF--N is 
the only place where all the ingredients are available for a single 
field.  Even there, our NICMOS map is neither as deep (to minimize 
surface brightness dimming losses) nor as sharp (to match WFPC2 angular 
resolution) as we would like.  Moreover, with only a single field 
covering a small cosmic volume, it is risky to generalize HDF--N 
results to the universe as a whole.  Nevertheless, several trends 
are apparent.  We find giant disk galaxies with prominent spiral 
structure, red bulges, and bars out to $z \sim 1.25$, and red, 
apparently mature giant ellipticals out to $z_{phot} \approx 1.8$.  
The latter probably formed the bulk of their stars at much higher 
redshift.  However, bluer early type galaxies are also found at 
$0.5 \simlt z \simlt 1.4$, suggesting that some field ellipticals
had extended star formation histories, in contrast with what has 
been observed in rich clusters.  The morphological peculiarities 
of most irregular HDF galaxies persist in the NICMOS images, suggesting 
that they are genuinely disturbed or immature objects.  Finally, 
there seem to be far fewer high luminosity galaxies of all types at 
$1.4 \simlt z < 2$ compared to lower redshifts, although this 
result must be treated with caution given the small volume of 
the HDF, its susceptibility to clustering variations, and the 
reliance on photometric redshifts.  With a revived NICMOS in 
Cycle 10 we can extend such work to other fields.  Better still, 
the infrared channel of WFC3, scheduled for installation on HST 
in 2003, will have a substantially wider field of view and smaller 
pixel scale, providing improved angular resolution compared to the 
undersampled NICMOS Camera~3.  Deeper and wider surveys will become 
enormously more efficient, permitting a much better census of the 
distant universe in the near infrared, and helping to pave the way 
for NGST.

\section{Acknowledgements}

I would like to thank my collaborators on the HDF/NICMOS GO 
program for their contributions to this project, and for allowing 
me to present results in advance of publication.  I especially 
thank Tamas Budav\'ari for deriving the photometric redshifts 
used here, and Adam Stanford for analysis and discussions about 
HDF elliptical galaxies, and for compiling the morphological 
classifications.  I also thank the organizers of this meeting for 
their hospitality, generous travel support, and patience editing 
these proceedings.  Support for this work was provided by 
NASA grant GO-07817.01-96A.

\vfill

\end{document}